\documentclass[twocolumn,amsmath,amssymb,prd]{revtex4}
\usepackage{slashed}

\def\al{\alpha}
\def\be{\beta}
\def\ga{\gamma}

\def\ep{\epsilon}

\def\et{\eta}
\def\th{\theta}

\def\la{\lambda}

\def\rh{\rho}

\def\si{\sigma}

\def\ta{\tau}

\def\ph{\phi}

\def\ch{\chi}

\def\om{\omega}
\def\Ga{\Gamma}
\def\De{\Delta}

\def\mn{{\mu\nu}}
\def\ni{\noindent}

\newcommand{\beq}{\begin{equation}}
\newcommand{\eeq}{\end{equation}}
\newcommand{\bea}{\begin{eqnarray}}
\newcommand{\eea}{\end{eqnarray}}
\newcommand{\bit}{\begin{itemize}}
\newcommand{\eit}{\end{itemize}}
\newcommand{\bM}{\begin{pmatrix}}
\newcommand{\eM}{\end{pmatrix}}

\def\Re{\hbox{Re}\,}

\newcommand{\BB}{\big}
\newcommand{\nn}{\nonumber}

\def\sF#1#2{{\textstyle{{#1}\over{#2}}\,}}

\def\gM#1{g_{\be\be}^{#1}}
\def\mn{{\mu\nu}}

\def\etal {{\it et al.}}
\def\ol#1{\overline{#1}}
\def\sl#1{\slashed{#1}} 
\newcommand{\W}{\widetilde}
\def\aof{(a^{(3)}_\text{of})}
\def\ring#1{{\mathaccent'27 #1}}

\def\vev#1{\langle {#1}\rangle}

\newcommand{\arit}{\ring{a}_\text{of}^{(3)}}
\newcommand{\wT}{\om_\oplus T_\oplus}

\begin{document}

\title{Limits on Lorentz and CPT violation from double beta decay}
\author{Jorge S. D\'iaz}
\affiliation{Physics Department, Indiana University, Bloomington, IN 47405, USA}
% IUHET 583

\begin{abstract}
Deviations from Lorentz and CPT invariance in the neutrino sector and their observable effects in double beta decay are studied.
For two-neutrino double beta decay, a spectral distortion and its properties are characterized for different isotopes.
Majorana couplings for Lorentz violation are studied and shown to trigger neutrinoless double beta decay even for negligible Majorana mass.
Existing data are used to obtain first limits of $5 \times 10^{-9}$ for 18 individual coefficients and attainable sensitivities in current and future experiments are presented.

\end{abstract}

\maketitle

\section{Introduction}

The development of Fermi's theory of beta decay \cite{Fermi1934} was rapidly followed by important ideas involving weak interactions. In 1935, Goeppert-Mayer proposed the possibility that two neutrons in a nucleus could simultaneously decay into two protons, two electrons, and two antineutrinos \cite{GoeppertMayer1935}, estimating that this rare decay would have a half life greater than $10^{17}$ years.
This two-neutrino double beta decay is a second-order weak process of the form
$(A,Z)\to(A,Z+2)+2e^-+2\bar\nu_e$
that has been observed in many nuclei \cite{PDG2012,DBDreviews}.
Other authors also proposed the possibility of another mode of double beta decay, in which the neutrino would be absent in the final state. This neutrinoless double beta decay $(A,Z)\to(A,Z+2)+2e^-$ could be possible if neutrinos are their own antiparticles \cite{0vbb}.
Today, there is a dedicated experimental program searching for neutrinoless double beta decay. 
A careful study of two-neutrino double beta decay is also performed by these experiments because it constitutes a background for the neutrinoless mode.
The high precision of many experiments has motivated the formulation of different modes of double beta decay so that experiments can also look for new physics through unconventional decay modes \cite{DBDreviews}.

In the present work, we propose to use experiments studying double beta decay as probes of Lorentz invariance.
The spontaneous breakdown of this spacetime symmetry is an interesting feature that can be accommodated by many candidate theories of quantum gravity, such as string theory ~\cite{SBS_LV}.
Even though direct studies of physics at quantum-gravity energies remain inaccessible for current experiments, we can use low-energy experiments as tools to explore suppressed signals of new physics at the Planck scale.
The effects of violations of Lorentz symmetry in beta decay have been characterized in Ref.~\cite{DKL}, which briefly describes the experimental signals of Lorentz violation in double-beta-decay experiments.
In this work, we assume that Lorentz-violating effects only affect neutrinos. 
The corresponding effects arising from Lorentz violation in the gauge and Higgs sectors are active topics of research \cite{LV_weak}, which shows the growing interest on testing fundamental symmetries using weak decays \cite{Severijns2013}.
Couplings with weak gravitational fields are discussed in Ref. \cite{KK0vbb} and CPT violation without breaking Lorentz invariance is presented in Ref. \cite{Barenb0vbb}.

The general framework that incorporates operators that break Lorentz invariance in the SM is the Standard-Model Extension (SME) \cite{SME}.
This effective field theory parametrizes generic deviations from Lorentz invariance in the form of coordinate-invariant terms in the action by contracting operators of conventional fields with controlling coefficients for Lorentz violation.
This construction guarantees invariance of the theory under observer transformations, whereas particle Lorentz symmetry is broken.
It should be noted that a subset of operators in the SME also break CPT symmetry \cite{CPTv}.
The development of the SME has led a worldwide experimental program searching for violations of Lorentz invariance, whose results are summarized in Ref. \cite{tables}. 

The study of the neutrino sector of the SME \cite{LVnu} has characterized the high sensitivity of neutrino-oscillation experiments.
The development of methodologies to perform systematic searches for Lorentz violation in these experiments \cite{KM_SB,DKM} has motivated several experimental searches using neutrinos and antineutrinos \cite{LV_DC,LV_IceCube,LV_LSND,LV_MiniBooNE,LV_MINOS,LV_SK}.
Additionally, the SME has been used to construct alternative models for neutrino oscillations that can accommodate the established data and also some of anomalous results reported by different experiments \cite{LVmodels}.
Some of these models based on the SME offer elegant and interesting solutions to neutrino anomalies.
More recently, the observation of very-high-energy neutrinos \cite{IceCube_PeVnus} has served to determine stringent constraints on CPT-even SME coefficients \cite{VHEnus}.

Even though the interferometric nature of neutrino oscillations makes them sensitive tools to search for new physics, the study of weak decays offers access to some operators that are unobservable using neutrino mixing.
Operators of arbitrary dimension in the theory can be studied using the methods introduced in Ref. \cite{KM2012}. 
Nonetheless, of particular interest are those whose observable effects escape detection through sensitive measurements such as neutrino oscillations and time of flight.
These so-called {\it countershaded} effects \cite{DKL,KT} arise due to oscillation-free operators of mass dimension three, whose effects are controlled by the four independent components of the coefficient denoted $\aof^\al$ \cite{KM2012}.

\section{Two-neutrino double beta decay}

The unconventional spinor solutions of the modified Dirac equation and the form of the neutrino phase space 
produce observable effects recently studied in the context of tritium decay, neutron decay, and
double beta decay \cite{DKL}. 
In this section, we describe a detailed presentation of the relevant experimental signature in the two-neutrino mode of double beta decay.

Denoting the 4-momentum of the two electrons and the two antineutrinos by $p_j^\al=(E_j,\pmb{p}_j)$ and 
$q_j^\al=(\om_j,\pmb{q}_j)$, respectively ($j=1,2$), the relevant matrix element for the two-neutrino mode of double beta decay is given by
\bea
i\mathcal{M} &=& iG_F^2V_{ud}^2 [\ol u(p_1)\ga^\mu(1-\ga_5)v(q_1)]
\nn\\
&&\times\,
[\ol u(p_2)\ga^\nu(1-\ga_5)v(q_2)]\,J_\mn
-(p_1\leftrightarrow p_2).
\label{M_2v}
\eea

\ni
The hadronic tensor $J_\mn$ corresponds to the product of two nuclear currents written in the impulse approximation \cite{DBDreviews}.
Following the same procedure as in the conventional two-neutrino double beta decay, including the implementation of the long-wave and closure approximation for the hadronic tensor \cite{DBDreviews}, we obtain
\beq
\sum_\text{spin}|\mathcal{M}|^2 =
64 G_F^4|V_{ud}|^4g_A^4\, 
(p_1\cdot p_2)(\W q_1\cdot \W q_2) |M^{2\nu}|^2,
\eeq

\ni
where the nuclear matrix element involves vector and axial couplings for Fermi and Gamow-Teller transitions in the form $g_A^2M^{2\nu}=g_V^2 M^{2\nu}_F-g_A^2M^{2\nu}_{GT}$ \cite{DBDreviews}.
The two antineutrinos appear with an effective 4-momentum $\W q^\al=(\om,\pmb{q}+\pmb{a}_\text{of}^{(3)}-\arit\,\hat{\pmb{q}})$,
where $\arit$ corresponds to the isotropic component of $\aof^\al$.
The decay rate is given by
\bea
d\Ga &=& \frac{1}{4}\int
\frac{d^3p_1}{(2\pi)^32E_1} \frac{d^3p_2}{(2\pi)^32E_2}
\frac{d^3q_1}{(2\pi)^32\om_1} \frac{d^3q_2}{(2\pi)^32\om_2}
\nn\\
&&\qquad\times\,
F(Z,E_1)F(Z,E_2)\;\textstyle{\sum}|\mathcal{M}|^2
\nn\\
&&\qquad\times\,
2\pi \delta(E_1+E_2+\om_1+\om_2-\De M) ,
\eea

\ni
where we have included the Fermi function to account for the Coulomb interaction of the two emitted electrons and the daughter nucleus of atomic number $Z$, and the symmetry factor for the two pairs of outgoing leptons.
Since the two antineutrinos are not measured in these type of experiments, the integration over all orientations leaves only the isotropic coefficient $\arit$ and the phase space takes the form $d^3q=4\pi(\om^2+2\om\,\arit)\,d\om$.
After a suitable change of integration variables and defining the sum of kinetic energies $K=T_1+T_2$ for the two electrons, we obtain the electron sum spectrum 
\bea
\frac{d\Ga}{dK} &=& C\BB(K^4+10K^3+40K^2+60K+30\BB)\,K
\nn\\
&&\times\,\BB[(K_0-K)^5+10\arit (K_0-K)^4\BB],
\eea

\ni
where $K_0$ is the maximum kinetic energy available in the decay.
We have written the energy in units of the electron mass and used the Primakoff-Rosen limit for nonrelativistic electrons \cite{PR1959,Konopinski_book, Doi, Haxton_Stephenson, Tomoda}.
The overall constant factor is given by $C=G_F^4g_A^4|V_{ud}|^4 |M^{2\nu}|^2F^2_\text{PR}(Z)m_e^{11}/7200\pi^7$, with $F_\text{PR}(Z)=2\pi\al Z/(1-e^{-2\pi\al Z})$.
We find that the isotropic coefficient $\arit$ produces a distortion of the conventional electron sum spectrum.
A similar effect is found for studies of the spectrum of neutron decay \cite{DKL}.
Since the Lorentz-violating modification of the spectrum appears with a well-defined energy dependence, a search for deviations from the conventional spectrum would allow studying the effects of the isotropic coefficient $\arit$.
In particular, the energy dependence of the modification allows determining the energy $K_m$ at which the effect of this coefficient is maximal and hence giving the best chance to observe its effect.
This is the energy at which the residual spectrum reaches its maximum.
Table \ref{Table:K_m} shows the value of this energy for some double-beta-decay emitters commonly studied.
It should be noticed that the coefficient $\arit$ also controls a source of CP violation in the neutrino sector that remains experimentally unexplored.

\begin{table}\begin{center}
\begin{tabular}{cc|cc}
\hline\hline
Isotope &   $K_m$ (MeV) &   Isotope &   $K_m$ (MeV) \\ \hline
$^{48}$Ca   &   1.98    &   $^{116}$Cd  &   1.20    \\
$^{76}$Ge   &   0.81    &   $^{130}$Te  &   1.05    \\
$^{82}$Se   &   1.30    &   $^{136}$Xe  &   1.02    \\
$^{96}$Zr   &   1.48    &   $^{150}$Nd  &   1.49    \\
$^{100}$Mo  &   1.32    &               \\
\hline\hline
\end{tabular} 
\caption{Sum of the kinetic energy of the two electrons at which the Lorentz-violating modification introduced by the coefficient $\arit$ is maximal. }
\label{Table:K_m}
\end{center} \end{table}

Experimental studies would require the search for deviations from the conventional spectrum and the energy listed in Table \ref{Table:K_m} for a given element should serve as a guide indicating the region of the spectrum of highest sensitivity.
We have also found that the anisotropic components of $\aof^\al$ are unobservable in this kind of experiments; therefore, studies of neutron and tritium decay appear as important complementary techniques \cite{DKL}.

\section{Neutrinoless double beta decay}

The presence of Lorentz violation in the neutrino sector modifies both the neutrino dispersion relation and propagator. 
This means that Lorentz-violating effects arise in two independent manners involving different types of coefficients. 
As discussed before, the modified neutrino dispersion relation alters the phase space, which for the neutrinoless mode introduces a modification to the so-called {\it neutrino potential} \cite{DBDreviews}.
It can be shown that current limits on the lifetime of neutrinoless mode of different isotopes are less sensitive to the relevant SME coefficients than tritium decay experiments \cite{DKL}. 
Additionally, the effects of Lorentz violation in the neutrino potential are weighted with the conventional nuclear matrix elements, which having magnitudes $\mathcal{O}(1)$ are unable to make these types of Lorentz-violating effects more noticeable.
For these reasons, we neglect the Lorentz-violating modifications of the nuclear matrix elements introduced by the coefficient $\arit$ and study in detail the effects of Lorentz-violating Majorana couplings that arise in the neutrino propagator.

Using the SME lagrangian \cite{KM2012} and neglecting Dirac couplings, the relevant propagator can be written at leading order  as 
\bea\label{S(Q)}
S(q)&=&\frac{1}{q^2}\BB[\sl{q}
-e^{(4)\la}_{Ma'\bar{a'}}\, q_\la
-if^{(4)\la}_{Ma'\bar{a'}}\, q_\la\ga_5
\nn\\
&&\qquad
-\sF{1}{2}g^{(4)\la\rh\si}_{Ma'\bar{a'}}\,\si_{\la\rh}q_\si 
-\sF{1}{2}H^{(3)\la\rh}_{Ma'\bar{a'}}\,\si_{\la\rh}\BB],
\eea 
where the coefficients $e^{(4)\la}_{Ma'\bar{a'}}, f^{(4)\la}_{Ma'\bar{a'}}, g^{(4)\la\rh\si}_{Ma'\bar{a'}}$, and $H^{(3)\la\rh}_{Ma'\bar{a'}}$ are Majorana couplings in the neutrino sector of the SME.
The indices $a'\bar{a'}$ indicate that the coefficients are written in the basis of neutrino eigenstates, with $a'=1,2,3$. 
The bar over the second index reveals the Majorana nature of these coefficients by connecting neutrino and antineutrino states.
The numbers in parentheses denote the mass dimension of the associated operator.
We have neglected the contribution from a possible Majorana mass in the propagator because we are interested in a pure Lorentz-violating mechanism for neutrinoless double beta decay.
Inspection shows that the coefficient $g^{(4)\la\rh\si}_{Ma'\bar{a'}}$ is the only one in the propagator \eqref{S(Q)} that couples to an operator that preserves charge conjugation.
This suggests that $g^{(4)\la\rh\si}_{Ma'\bar{a'}}$ is the only relevant coefficient in this decay mode because neutrinoless double beta decay requires the neutrino and the antineutrino to be the same particle.

Direct calculation shows that due to the Dirac-matrix structure, the transition matrix generated by the scalar $e^{(4)\la}_{Ma'\bar{a'}}$ and pseudoscalar $f^{(4)\la}_{Ma'\bar{a'}}$ couplings as well as the tensor coupling $H^{(3)\la\rh}_{Ma'\bar{a'}}$ is symmetric under the interchange of the two emitted electrons.
For this reason, the total transition matrix vanishes due to its antisymmetry under the interchange of identical fermions. 
In contrast, the coefficient $g^{(4)\la\rh\si}_{Ma'\bar{a'}}$ produces a transition matrix that is antisymmetric under the interchange of the two electrons.
This result verifies that Lorentz-violating neutrinoless double beta decay only depends on the coefficient $g^{(4)\la\rh\si}_{Ma'\bar{a'}}$, which also controls CPT violation.
The result above can be shown to be valid for operators of arbitrary dimension; nevertheless, in this work we only describe the effects of operators of lowest dimension, in this case $d=4$ \cite{KM2012}.

Integration over the neutrino energy shows that the effects of $g^{(4)\la\rh0}_{Ma'\bar{a'}}$ are unobservable; 
hence, only $g^{(4)\la\rh k}_{Ma'\bar{a'}}$ is relevant.
The Dirac matrix $\si_{\la\rh}$ introduces a coupling between the components of the coefficient $g^{(4)\la\rh k}_{Ma'\bar{a'}}$ and the phase space of the two emitted electrons, which leads to unique electron angular correlations. 
Other unconventional terms appear due to the coupling between $g^{(4)\la\rh k}_{Ma'\bar{a'}}$ and the components of the nuclear currents. 
It can be shown that Dirac-matrix structure can be written as the sum of symmetric, antisymmetric, and mixed products of Dirac matrices. The symmetric piece leads to the conventional Fermi and Gamow-Teller nuclear matrix elements, while the antisymmetric part produces no effects because the product of nuclear currents is symmetric \cite{DBDreviews}.
The mixed term leads to unconventional forms for the nuclear matrix elements.
For a detailed study a careful analysis of the nuclear matrix elements coupled to the $g^{(4)\la\rh k}_{Ma'\bar{a'}}$ coefficient is needed, which goes beyond the scope of this work.
In what follows we will focus on the effects arising from the coupling between $g^{(4)\la\rh k}_{Ma'\bar{a'}}$ and the phase space of the two outgoing electrons, in which case the nuclear matrix elements remain unchanged.

The implementation of the results discussed above leads to the transition amplitude
\bea
i\mathcal{M} &=&
iG_F^2V_{ud}^2 \,\vev{g^{\la\rh k}_{M}} \hat r_k
\,\ol{u}(p_1)\ga^\mu\si_{\la\rh}\ga^\nu(1+\ga_5)u^C(p_2)
\nn\\
&&\qquad\times\,
\frac{H(\pmb{r},\De)\,J_\mn}{4\pi R^2},
\eea

\ni
where $H(\pmb{r},\De)$ is the conventional neutrino potential from the integration over the neutrino momentum \cite{DBDreviews}, the nuclear radius is given by $R=6.1\,\times A^{1/3}$ GeV$^{-1}$, and $\hat r$ is the direction of emission of the virtual neutrino.
The coefficients for Lorentz violation appear weighted by the electron elements of Pontecorvo-Maki-Nakagawa-Sakata matrix in the form of an effective coefficient
\beq\label{<g>}
\vev{g^{\la\rh k}_{\be\be}} = \sum_{a'}U_{a'e}^2\, g^{(4)\la\rh k}_{Ma'\bar{a'}},
\eeq
similar to the conventional effective Majorana-mass parameter $\vev{m_{\be\be}}$ \cite{DBDreviews}.
From direct calculation we obtain
\bea\label{M_0v}
\sum_\text{spin} |\mathcal{M}|^2 &=&
\frac{G_F^4|V_{ud}|^4g_A^4}{8\pi^2R^4}\,|M^{0\nu}|^2
\,|\gM{\la\rh}|^2 
\BB[ 
(p_1\cdot p_2)\,\et^{\la\la}\et^{\rh\rh} 
\nn\\
&&\qquad
- 2p_1^\la p_2^\la \et^{\rh\rh}
- 2p_1^\rh p_2^\rh \et^{\la\la}\BB],
\eea

\ni
where $M^{0\nu}$ is the conventional nuclear matrix element for neutrinoless double beta decay \cite{DBDreviews}.
In this expression we have written the vector $\hat r$ in the Sun-centered frame \cite{SunFrame} and averaged over all possible orientations, which produces the effective coefficient
\beq\label{g_M}
|\gM{\la\rh}|^2=\frac{1}{3}\BB(|\vev{g^{\la\rh X}_{\be\be}}|^2+|\vev{g^{\la\rh Y}_{\be\be}}|^2+|\vev{g^{\la\rh Z}_{\be\be}}|^2\BB).
\eeq

\ni
It must be noted that in the transition amplitude \eqref{M_0v} there is no sum over repeated indices. 
Instead, this expression is valid individually for the six possible pairs of indices $\la\rh$.
The decay rate is given by
\bea
d\Ga &=&
\frac{1}{2}\int
\frac{d^3p_1}{(2\pi)^32E_1}\frac{d^3p_2}{(2\pi)^32E_2}
\,F(Z,E_1)F(Z,E_2)\,
\nn\\
&&\qquad\times\,
\;\textstyle{\sum}|\mathcal{M}|^2
\,2\pi \delta(E_1+E_2+\De M),
\label{dGa_0v}
\eea

\ni
where Lorentz-violating effects only appear at the level of the transition amplitude because the physics of the two outgoing electrons is conventional.

\subsection{Angular correlations}

For experiments with tracking systems that allow the determination of the direction of the two emitted electrons it is useful to identify the way the angular correlation of the electrons gets modified.
In the Sun-centered frame \cite{SunFrame} the momentum of each electron satisfies
\beq
\frac{\pmb{p}}{E} = \be
\bM \cos\wT & -\sin\wT & 0 \\ \sin\wT & \cos\wT & 0 \\ 0&0&1\eM
\bM N^X \\ N^Y \\ N^Z\eM,
\eeq

\ni
where $\be$ is the speed of the electron and $\om_\oplus\simeq2\pi/(\text{23 h 56 min})$ is the sidereal frequency that accounts for the rotation of the Earth. 
The directional factors $N^X, N^Y, N^Z$ indicate the direction of each electron in the Sun-centered frame during the 2000 vernal equinox, which defines $T_\oplus=0$ \cite{SunFrame}.
Using polar coordinates in the laboratory frame with the $z$ axis directed towards the zenith, the $x$ axis pointing south, and the $y$ axis pointing east, the directional factors are functions of $(\th,\ph)$ and the colatitude of the laboratory $\ch$ \cite{DKM}.
Since we will integrate over the azimuthal angles $\ph_j$, we only keep the elements with azimuthal symmetry, in which case the momentum components of each electron $(j=1,2)$ become
\bea
p^X_{j} &=& \cos\wT\,\sin\ch\cos\th_{j},
\nn\\
p^Y_{j} &=& \sin\wT\,\sin\ch\cos\th_{j},
\nn\\
p^Z_{j} &=& \cos\ch\cos\th_{j}.
\label{p^J}
\eea

\ni
Using these expressions, we can write the angular distribution as
\beq
\rh_{\la\rh}(\pmb{p}_1,\pmb{p}_2)
= 1-k_{\la\rh}\,\be_1\be_2\cos\th_1\cos\th_2,
\eeq

\ni
with the factor $k_{\la\rh}$ given for each component in the form
\bea
&& k_{TX}= k_{YZ} = 2\sin^2\ch\cos^2\wT-1 , \nn\\
&& k_{TY}= k_{XZ} = 2\sin^2\ch\sin^2\wT-1 , \nn\\
&& k_{TZ}= k_{XY}= \cos2\ch.
\eea

\ni
We have found that all the factors $k_{\la\rh}$ depend on the location of the experiment and four of them change with sidereal time.
For comparison, in neutrinoless double beta decay triggered by a Majorana mass this factor is $k=1$ \cite{DBDreviews,Primakoff1952}.
In order to construct a useful observable we define the quantity 
\beq\label{K_ab}
K_{\la\rh}=k_{\la\rh}\,k_\th(Z),
\eeq

\ni
which includes the factors $k_{\la\rh}$ obtained above and also the parameter $k_\th(Z)$ that accounts for the integration over the electron energies and depends on the isotope used through the atomic number $Z$ and the $Q$ value.
These parameters must be determined by numerical integration of the decay rate \eqref{dGa_0v} over the allowed electron energies.
Table \ref{Table:k} shows the value of this parameter for some double beta decays.
\begin{table}%[ht] 
\begin{center}
\begin{tabular}{ccc|ccc}
\hline\hline
Process && $k_\th$ & Process && $k_\th$   \\ \hline
$^{48}$Ca$\to^{48}$Ti   &&  0.93    &   $^{116}$Cd$\to^{116}$Sn &&  0.87    \\
$^{76}$Ge$\to^{76}$Se       &&  0.81    &   $^{136}$Xe$\to^{136}$Ba     &&  0.84    \\
$^{82}$Se$\to^{82}$Kr   &&  0.88    &   $^{130}$Te$\to^{130}$Xe &&  0.85    \\
$^{96}$Zr$\to^{96}$Mo   &&  0.90    &   $^{150}$Nd$\to^{150}$Sm &&  0.89    \\
$^{100}$Mo$\to^{100}$Ru &&  0.88    &       &&      \\
\hline\hline
\end{tabular} 
\caption{Relevant parameters for some double beta decays.}
\label{Table:k}
\end{center} \end{table}
%\ni
The definition of the factor \eqref{K_ab} can be used to write
\beq
\frac{d\Ga}{dx_1dx_2}=\frac{\Ga}{4}\BB(1-K_{\la\rh} x_1x_2\BB),
\eeq

\ni
with $\Ga$ the total decay width and $x_j=\cos\th_j$. 
In the absence of Lorentz violation, the local coordinates can be taken with any orientation. 
In our case we do not have this freedom because the coordinates of the laboratory frame are defined according to the rotations used to write the two electron momenta in the Sun-centered frame \eqref{p^J}; therefore, the integration over the two electron orientations must be performed carefully. 
Let us define the number of events $N^-$ ($N^+$) emitted with a relative angle smaller (greater) than $90^\circ$ in the form
\bea
N^- &=& 
\int_{x_\text{min}}^{x_\text{max}}dx_1
\int_{-\sqrt{1-x_1^2}}^{x_1} dx_2
\bigg(\frac{d\Ga}{dx_1dx_2}\bigg),
\nn\\
N^+ &=& 
\int_{x_\text{min}}^{x_\text{max}}dx_1
\int_{-x_1}^{-\sqrt{1-x_1^2}} dx_2
\bigg(\frac{d\Ga}{dx_1dx_2}\bigg).
\label{N_pm}
\eea

\ni
The forward-backward asymmetry of the decay distribution can be constructed by properly choosing the range $[x_\text{min},x_\text{max}]$.
Given the form of the angular distribution, if the integration range is too symmetric then the terms of interest cancel. For this reason we take the range $[-\sF{1}{\sqrt2},1]$, which gives the asymmetry
\beq
\mathcal{A} = \frac{N^+-N^-}{N^++N^-} 
= \frac{K_{\la\rh}}{4}-\frac{3\pi}{2}-1.
\eeq

\ni
This asymmetry corresponds to the counting of all the events between $\th=0^\circ$ measured from the vertical at the laboratory frame and $\th=135^\circ$.
Notice that the above asymmetry depends on the location of the experiment and can also oscillate with sidereal time, in which case the amplitude of the oscillation is independent of the size of the coefficient for Lorentz violation.
The asymmetry in the conventional Majorana-mass driven decay is constant and only depends on the element used in the form $\mathcal{A}_0(Z)=k_\th(Z)/2$.
We find that the asymmetry defined above allows a clear separation of the effects due to $|\gM{\la\rh}|$ from the conventional neutrinoless double beta decay triggered by a Majorana-mass parameter.

\subsection{Half-life measurements}

The complete integration of the decay rate \eqref{dGa_0v} allows identifying the signature of the effective coefficients $|\gM{\la\rh}|$ in the  half life of neutrinoless double beta decay.
The decay constant can be written in the conventional form as the product of a phase-space factor $G^{0\nu}(Z,Q)$, that depends on the double beta emitter; the nuclear matrix element; and a particle physics quantity in the form
\beq\label{T0v}
(T^{0\nu}_{1/2})^{-1}= \Ga =
G^{0\nu}(Z,Q)\,|M^{0\nu}|^2\,\frac{|\gM{\la\rh}|^2}{4R^2}.
\eeq

\ni
Comparing this expression with the conventional form of the decay constant \cite{DBDreviews} we find that effective coefficients for CPT-odd Lorentz violation $|\gM{\la\rh}|$ play the role of the Majorana mass parameter $|\vev{m_{\be\be}}|$ in the form \cite{DKL}
\beq\label{m2g}
|\vev{m_{\be\be}}| \to \frac{|\gM{\la\rh}|}{2R}.
\eeq

\ni
Since neutrinoless double beta decay remains unobserved to date, limits on the half life of this decay mode in different isotopes can be used to set upper bounds on the effective coefficients $|\gM{\la\rh}|$.
Table \ref{Table:g_M} presents estimated limits on these coefficients based on published results by different experiments as well as the projected sensitivity of several experiments under construction or in the process of upgrade.

\begin{table}%[ht] 
\begin{center}
\begin{tabular}{cccc}
\hline\hline
Process & estimated upper limit  & experiment & Ref. \\ \hline
$^{48}$Ca$\to^{48}$Ti       &    $1\times 10^{-8}$  &   IGEX    &    \cite{IGEX}    \\
$^{48}$Ca$\to^{48}$Ti       &    $4\times 10^{-7}$  &   ELEGANT VI  &    \cite{ELEGANT} \\
$^{48}$Ca$\to^{48}$Ti       &    $3\times 10^{-7}$  &    NEMO-3     &    \cite{NEMO3(Ca48)} \\
$^{76}$Ge$\to^{76}$Se       &    $4\times 10^{-9}$  &    GERDA  &    \cite{GERDA}   \\
$^{82}$Se$\to^{82}$Kr       &    $3\times 10^{-8}$  &    NEMO-3 &    \cite{SuperNEMO}   \\
$^{96}$Zr$\to^{96}$Mo       &    $2\times 10^{-7}$  &    NEMO-3 &    \cite{NEMO3(Zr96)} \\
$^{100}$Mo$\to^{100}$Ru     &    $1\times 10^{-8}$  &    NEMO-3 &    \cite{SuperNEMO}   \\
$^{116}$Cd$\to^{116}$Sn     &    $2\times 10^{-8}$  &    Solotvina  &    \cite{Solotvina}   \\
$^{130}$Te$\to^{130}$Xe     &    $9\times 10^{-9}$  &    CUORICINO  &    \cite{CUORICINO}   \\
$^{136}$Xe$\to^{136}$Ba     &    $5\times 10^{-9}$  &    EXO-200    &    \cite{EXO(PRL2012)}    \\
$^{136}$Xe$\to^{136}$Ba     &    $3\times 10^{-9}$  &   KamLAND-Zen &    \cite{KamLANDZ(PRL2013)}   \\
$^{150}$Nd$\to^{150}$Sm     &    $8\times 10^{-8}$  &    NEMO-3 &    \cite{NEMO3(Nd150)}    \\
\hline\hline                            
$^{48}$Ca$\to^{48}$Ti       &   $4\times 10^{-9}$   &   CANDLES &   \cite{CANDLES}  \\
$^{76}$Ge$\to^{76}$Se   &   $4\times 10^{-10}$  &   GERDA Phase II  &   \cite{GERDA}    \\
$^{76}$Ge$\to^{76}$Se   &   $4\times 10^{-10}$  &   MAJORANA    &   \cite{MAJORANA} \\
$^{82}$Se$\to^{82}$Kr   &   $6\times 10^{-9}$   &   SuperNEMO   &   \cite{SuperNEMO}    \\
$^{130}$Te$\to^{130}$Xe &   $5\times 10^{-9}$   &   CUORE-0   &   \cite{CUORE0}    \\
$^{130}$Te$\to^{130}$Xe &   $1\times 10^{-9}$   &   CUORE   &   \cite{CUORE}    \\
$^{130}$Te$\to^{130}$Xe &   $8\times 10^{-10}$  &   SNO+    &   \cite{SNO+} \\
$^{136}$Xe$\to^{136}$Ba     &   $4\times 10^{-10}$   &   nEXO    &   \cite{nEXO} \\
$^{136}$Xe$\to^{136}$Ba     &   $1\times 10^{-9}$   &   NEXT    &   \cite{NEXT} \\
\hline\hline
\end{tabular} 
\caption{Conservative upper limits on the effective coefficients $|\gM{\la\rh}|$ from the corresponding upper bound on the effective Majorana mass at the 90\% C.L. in different experiments.
Lower rows show attainable limits based on the expected sensitivities in future experiments.}
\label{Table:g_M}
\end{center} \end{table}

The limits presented in Table \ref{Table:g_M} are conservative in the sense that the lowest value of the corresponding nuclear matrix element has been used. 
Limits two to three times better can be obtained when using the largest value of the nuclear matrix elements.
Since the definition of $|\gM{\la\rh}|$ involves the sum of three positive quantities, 
we can take each of the components $|\vev{g^{\la\rh K}_{\be\be}}|$ in the definition \eqref{g_M} to be nonzero at the time. 
From the values in Table \ref{Table:g_M} we can write 90\% C.L. upper limits in 18 effective coefficients in the form
\beq\label{|<g>|}
|\vev{g^{\la\rh K}_{\be\be}}| < 5 \times 10^{-9},
\eeq
with $\la\rh=TX, TY, TZ, XY, XZ, YZ,$ and $K=X,Y,Z$.
The definition of the effective coefficients $\vev{g^{\la\rh K}_{\be\be}}$ \eqref{<g>} shows that they are linear combinations of the eigenvalues of the original coefficients $g_M^{(4)\la\rh\si}$ in the action.
It should be noticed that different linear combinations of these original coefficients can trigger neutrino-antineutrino oscillations \cite{LVnu,DKM}, whose effects have been recently studied using accelerator neutrinos \cite{RebelMufson} and reactor antineutrinos \cite{LV_DC2}. 
The interferometric nature of oscillations between neutrinos and antineutrinos \cite{DKM} make them sensitive probes of the Majorana couplings in the SME, which appear as combinations of the original coefficients in the form 
$\W g^{\nu\si}_{a\bar b} = g^{T\nu\si}_{a\bar b}+\sF{i}{2}\ep^{T\nu}_{\quad\ga\rh}\,g^{\ga\rh\si}_{a\bar b}$ ($a=e,\mu,\ta$,
$\bar b=\bar e,\bar\mu,\bar\ta$) \cite{LVnu}.
The experimental signature in neutrinoless double beta decay allows constraining independent combinations of these coefficients \cite{Disentangling}, which provides complementary tests of Lorentz invariance.

Even though neutrinoless double beta decay can access different combinations of coefficients from those of oscillations, under mild assumptions these combinations can be related. 
Below we present the relationships between the effective coefficient $|\gM{\la\rh}|$ observable in neutrinoless double beta decay and some components of the coefficients measured in oscillations, which are obtained by keeping a single coefficient while setting the others to zero \footnote{The Pontecorvo-Maki-Nakagawa-Sakata matrix is used to diagonalize the $3\times3$ Majorana block at leading order. The values $\sin^22\th_{12}=0.855,\,\sin^22\th_{13}=0.099,\,\sin^22\th_{23}=1.000$ have been used.}.
This procedure is widely used in the literature and it could hide some effects due to fortuitous cancellations between different coefficients; nonetheless, this method provides meaningful information for comparison with other experiments. 
The relations between coefficients are:
\begin{align}
\BB|\W g^{JK}_{e\bar e}\BB| &< \;\;3.9\, |\vev{\gM{TJ}}|, \quad &
\BB|\Re\W g^{JK}_{e\bar\mu}\BB| &< 3.2\, |\vev{\gM{TJ}}|, \nn\\
\BB|\W g^{JK}_{\mu\bar\mu}\BB| &< 10.8\, |\vev{\gM{TJ}}|, \quad &
\BB|\Re\W g^{JK}_{e\bar\ta}\BB| &< 5.3\, |\vev{\gM{TJ}}|, \nn\\
\BB|\W g^{JK}_{\ta\bar\ta}\BB| &< 28.7\, |\vev{\gM{TJ}}|, \quad &
\BB|\Re\W g^{JK}_{\mu\bar\ta}\BB| &< 8.9\, |\vev{\gM{TJ}}|, 
\end{align}

\ni
with $J,K=X,Y,Z$.
The constraint \eqref{|<g>|} leads to the following limits for 54 coefficients in flavor space:
\begin{align}
\BB|\W g^{JK}_{e\bar e}\BB| &< 1\times10^{-8} , \quad &
\BB|\Re\W g^{JK}_{e\bar\mu}\BB| &< 1\times10^{-8}, \nn\\
\BB|\W g^{JK}_{\mu\bar\mu}\BB| &< 3\times10^{-8}, \quad &
\BB|\Re\W g^{JK}_{e\bar\ta}\BB| &< 2\times10^{-8}, \nn\\
\BB|\W g^{JK}_{\ta\bar\ta}\BB| &< 9\times10^{-8}, \quad &
\BB|\Re\W g^{JK}_{\mu\bar\ta}\BB| &< 3\times10^{-8}.
\end{align}

\section{Conclusion}

In this work we have presented the effects of deviations from exact Lorentz invariance in  experiments studying double beta decay in the context of the Standard-Model Extension. 
Observable signatures include a modification of the electron sum spectrum in the two-neutrino mode of the decay, which can be studied by searching for departures from the conventional spectrum. 
The countershaded coefficient responsible for this effect controls Lorentz and CPT violation in the neutrino sector and also controls a new source of CP violation.
We also find that neutrinoless double beta decay could occur even if the Majorana neutrino mass is negligible; nevertheless, interference between the mass mechanism and the Lorentz-violating effect could also appear \cite{DKL,DBD_interf}. 
The corresponding angular correlations for the two emitted electrons have been determined for the relevant SME coefficients.
Although there are other mechanisms that can trigger neutrinoless double beta decay without the conventional Majorana mass \cite{DBDreviews,0vbb_m0}, the one presented here does not require new particles or forces.

In addition to limits on the Majorana mass parameter, the identification \eqref{m2g} will allow experiments to derive upper limits on coefficients for Lorentz violation from the lower bounds on the lifetime of neutrinoless double beta decay.
Notice that if neutrinoless double beta decay is observed in the future, the Lorentz-violating unique signature that would allow separating the effects from a Majorana-mass mechanism is the dependence of the particle physics parameter in the half life \eqref{T0v} on the nuclear radius $R$ of the isotope used.

The absence of compelling positive signals of neutrinoless double beta decay in numerous experiments is used to determine the first limits on some components of the relevant SME coefficient up to the $10^{-9}$ level.
Expected sensitivity of upcoming experiments could improve these limits by at least one order of magnitude.

\section*{Acknowledgments}
The author thanks Alan Kosteleck\'y for many valuable discussions.
This work was supported in part by the Department of Energy under grant DE-FG02-13ER42002 and by the Indiana University Center for Spacetime Symmetries.


\begin{thebibliography}{9}

\bibitem{Fermi1934} 
E.~Fermi,
% An attempt of a theory of beta radiation
Z.\ Phys.\  {\bf 88}, 161 (1934).

\bibitem{GoeppertMayer1935} 
M.~Goeppert-Mayer,
% Double beta-disintegration
Phys.\ Rev.\  {\bf 48}, 512 (1935).

\bibitem{PDG2012}
J. Beringer {\it et al.} 
(Particle Data Group), 
Phys.\ Rev.\ D {\bf 86}, 010001 (2012).

\bibitem{DBDreviews}
For reviews see: 
S.R.~Elliott and P.~Vogel,
% DOUBLE BETA DECAY
Annu. Rev. Nucl. Part. Sci. {\bf52}, 115 (2002);
F.T.~Avignone, III \etal, 
% Double Beta Decay, Majorana Neutrinos, and Neutrino Mass,
Rev.\ Mod.\ Phys.\  {\bf 80}, 481 (2008);
S.M.~Bilenky,
% Neutrinoless double beta-decay,
Phys.\ Part.\ Nucl.\  {\bf 41}, 690 (2010);
W.~Rodejohann,
% Neutrino-less Double Beta Decay and Particle Physics,
Int.\ J.\ Mod.\ Phys.\ E {\bf 20}, 1833 (2011).

\bibitem{0vbb} 
E.~Majorana,
% Theory of the Symmetry of Electrons and Positrons,
Nuovo Cim.\  {\bf 14}, 171 (1937);
G.~Racah,
% On the symmetry of particle and antiparticle,
Nuovo Cim.\  {\bf 14}, 322 (1937);
W.H.~Furry,
%``On transition probabilities in double beta-disintegration,
Phys.\ Rev.\  {\bf 56}, 1184 (1939).

\bibitem{SBS_LV}
V.A.\ Kosteleck\'y and S. Samuel,
Phys.\ Rev.\ D {\bf 39}, 683 (1989);
V.A.\ Kosteleck\'y and R.\ Potting,
Nucl. Phys. B {\bf 359}, 545 (1991);
Phys.\ Rev.\ D {\bf 51}, 3923 (1995).

\bibitem{DKL}
J.S. D\'iaz {\it et al.}, 
Phys.\ Rev.\ D {\bf88}, 071902(R) (2013).

\bibitem{LV_weak}
J.P.~Noordmans \etal,
% Lorentz violation in neutron and allowed nuclear beta decay,
Phys.\ Rev.\ C {\bf 87}, 055502 (2013);
% Limits on Lorentz violation from forbidden beta decays,
Phys.\ Rev.\ Lett.\  {\bf 111}, 171601 (2013);
H.W.\ Wilschut \etal,
% A new approach to test Lorentz invariance,
Ann. Phys. (Berlin) {\bf525}, 653 (2013);
S.E.\ M\"uller \etal,
% First test of Lorentz invariance in the weak decay of polarized nuclei
Phys.\ Rev.\ D {\bf88}, 071901(R) (2013);
B.~Altschul,
% Contributions to Pion Decay from Lorentz Violation in the Weak Sector,
Phys.\ Rev.\ D {\bf88}, 076015 (2013).

\bibitem{Severijns2013} 
N.~Severijns and O.~Naviliat-Cuncic,
% Structure and symmetries of the weak interaction in nuclear beta decay,
Phys.\ Scripta {\bf T152}, 014018 (2013).

\bibitem{KK0vbb} 
H.~V.~Klapdor-Kleingrothaus \etal,
% Test of special relativity and equivalence principle from neutrinoless double beta decay
Eur.\ Phys.\ J.\ A {\bf 5}, 3 (1999).

\bibitem{Barenb0vbb} 
G.~Barenboim \etal,
% CPT violation and the nature of neutrinos,''
Phys.\ Lett.\ B {\bf 537}, 227 (2002).

\bibitem{SME}
D.\ Colladay and V.A.\ Kosteleck\'y,
Phys.\ Rev.\ D {\bf 55}, 6760 (1997);
Phys.\ Rev.\ D {\bf 58}, 116002 (1998);
V.A.\ Kosteleck\'y,
Phys.\ Rev.\ D {\bf 69}, 105009 (2004).

\bibitem{CPTv}
O.W.\ Greenberg,
Phys.\ Rev.\ Lett.\ {\bf 89}, 231602 (2002);
arXiv:1105.0927.

\bibitem{tables}
{\it Data Tables for Lorentz and CPT Violation},
V.A.\ Kosteleck\'y and N.\ Russell,
Rev.\ Mod.\ Phys.\  {\bf 83}, 11 (2011) 
[2013 edition arXiv:0801.0287v6].

\bibitem{LVnu}
V.A.\ Kosteleck\'y and M.\ Mewes,
Phys.\ Rev.\ D {\bf 69}, 016005 (2004).

\bibitem{KM_SB}
V.A.\ Kosteleck\'y and M.\ Mewes,
% Lorentz violation and short-baseline neutrino experiments
Phys.\ Rev.\ D {\bf 70}, 076002(R) (2004).

\bibitem{DKM}
J.S.\ D\'iaz {\it et al.}, 
Phys. Rev. D {\bf 80}, 076007 (2009).

\bibitem{LV_DC}
Y.\ Abe {\it et al.} [Double Chooz Collaboration],
Phys.\ Rev.\ D {\bf 86}, 112009 (2012).

\bibitem{LV_IceCube}
R.\ Abbasi {\it et al.}  [IceCube Collaboration],
Phys.\ Rev.\ D {\bf 82}, 112003 (2010).

\bibitem{LV_LSND}
L.B.\ Auerbach {\it et al.} [LSND Collaboration],
Phys.\ Rev.\ D {\bf 72}, 076004 (2005).

\bibitem{LV_MiniBooNE}
A.A.~Aguilar-Arevalo {\it et al.} [MiniBooNE Collaboration],
Phys.\ Lett.\ B {\bf 718}, 1303 (2013);
T.\ Katori [MiniBooNE Collaboration],
Mod.\ Phys.\ Lett.\ A {\bf 27}, 1230024 (2012).

\bibitem{LV_MINOS}
P.\ Adamson {\it et al.} [MINOS Collaboration],
Phys.\ Rev.\ Lett.\ {\bf 101}, 151601 (2008);
Phys.\ Rev.\ Lett.\ {\bf 105}, 151601 (2010);
Phys.\ Rev.\ D {\bf 85}, 031101 (2012).

\bibitem{LV_SK}
T.~Akiri [Super-Kamiokande Collaboration],
%``Sensitivity of atmospheric neutrinos in Super-Kamiokande to Lorentz violation,''
arXiv:1308.2210.

\bibitem{LVmodels}
V.A.\ Kosteleck\'y and M.\ Mewes,
Phys.\ Rev.\ D {\bf 70}, 031902 (2004);
T.\ Katori {\it et al.}, 
Phys.\ Rev.\ D {\bf 74}, 105009 (2006);
V.\ Barger {\it et al.}, Phys.\ Lett.\ B {\bf 653}, 267 (2007);
Phys.\ Rev.\ D {\bf 84}, 056014 (2011);
J.S.\ D\'\i az and V.A.\ Kosteleck\'y, 
Phys.\ Lett.\ B {\bf 700}, 25 (2011);
Phys.\ Rev.\ D {\bf 85}, 016013 (2012);
S.-J.~Rong and Q.-Y.~Liu,
% The perturbed puma model,
Chin.\ Phys.\ Lett.\  {\bf 29}, 041402 (2012).

\bibitem{IceCube_PeVnus}
M.G. Aartsen \etal\ [IceCube Collaboration], 
Phys.\ Rev.\ Lett. {\bf 111}, 021103 (2013);
Science 342, 1242856 (2013).

\bibitem{VHEnus}
J.S.\ D\'\i az \etal, 
% Testing Relativity with High-Energy Astrophysical Neutrinos
Phys.\ Rev.\ D {\bf 89}, 043005 (2014).

\bibitem{KM2012}
V.A.\ Kosteleck\'y and M.\ Mewes,
Phys.\ Rev.\ D {\bf 85}, 096005 (2012);
% Fermions with Lorentz-violating operators of arbitrary dimension,
Phys. Rev. D {\bf 88}, 096006 (2013).

\bibitem{KT} 
V.A.~Kosteleck\'y and J.~Tasson,
% Prospects for Large Relativity Violations in Matter-Gravity Couplings
Phys.\ Rev.\ Lett.\  {\bf 102}, 010402 (2009);
V.A.~Kosteleck\'y and J.~Tasson,
% Matter-gravity couplings and Lorentz violation
Phys.\ Rev.\ D {\bf 83}, 016013 (2011).

\bibitem{PR1959}
 H. Primakoff and S.P. Rosen,
 % Double beta decay
 Rep. Prog. Phys. {\bf 22}, 121 (1959).

\bibitem{Konopinski_book}
E.J.~Konopinski, 
{\it Theory of Beta Radioactivity} 
(Oxford University Press, London, England, 1966).
 
\bibitem{Doi}
M.\ Doi \etal, 
% Neutrino mass, the RH interaction and Double Beta Decay
Prog. Theor. Phys. {\bf66}, 1739 (1981);
Prog. Theor. Phys. 69, 602 (1983).

\bibitem{Haxton_Stephenson}
W.C.\ Haxton and G.J.\ Stephenson Jr., 
Prog. Part. Nucl. Phys. {\bf12}, 409 (1984).

\bibitem{Tomoda}
T.\ Tomoda, 
% Double beta decay
Rep. Prog. Phys. {\bf54}, 53 (1991).

\bibitem{SunFrame} 
R.~Bluhm {\it et al.},
Phys.\ Rev.\ Lett.\  {\bf 88}, 090801 (2002);
Phys.\ Rev.\ D {\bf 68}, 125008 (2003);
V.A.~Kosteleck\'y and M.~Mewes,
Phys.\ Rev.\ D {\bf 66}, 056005 (2002).

\bibitem{Primakoff1952}
H. Primakoff,
% Angular Correlation of Electrons in Double Beta-Decay
Phys.\ Rev. {\bf 85}, 888 (1952).
  
\bibitem{IGEX}
C.E. Aalseth {\it et al.}  [IGEX Collaboration],
% IGEX 76Ge neutrinoless double-beta decay experiment: Prospects for next generation experiments
Phys.\ Rev.\ D {\bf 65}, 092007 (2002).

\bibitem{ELEGANT}
I. Ogawa {\it et al.},
% Search for neutrino-less double beta decay of 48Ca by CaF2 scintillator 
Nucl.\ Phys.\ A {\bf 730}, 215 (2004).

\bibitem{NEMO3(Ca48)} 
R.L.\ Flack [NEMO-3 Collaboration],
J. Phys. Conf. Ser. {\bf136}, 022032 (2008).

\bibitem{GERDA} 
M.~Agostini {\it et al.}  [GERDA Collaboration],
%  Results on neutrinoless double beta decay of 76Ge from GERDA Phase I,
Phys.\ Rev.\ Lett.\  {\bf 111}, 122503 (2013).

\bibitem{SuperNEMO} 
L.~Simard [NEMO-3 and SuperNEMO Collaboration],
% The NEMO-3 experiment and the SuperNEMO project
Prog.\ Part.\ Nucl.\ Phys.\  {\bf 64}, 270 (2010).

\bibitem{NEMO3(Zr96)}
J.~Argyriades {\it et al.}  [NEMO-3 Collaboration],
% Measurement of the two neutrino double beta decay half-life of Zr-96 with the NEMO-3 detector,
Nucl.\ Phys.\ A {\bf 847}, 168 (2010).

\bibitem{Solotvina} 
F.A.~ Danevich {\it et al.},
% Search for 2? decay of cadmium and tungsten isotopes: Final results of the Solotvina experiment
Phys.\ Rev.\ C {\bf 78}, 035502 (2008).

\bibitem{CUORICINO} 
E.~Andreotti \etal~ [CUORICINO Collaboration],
% 130Te Neutrinoless Double-Beta Decay with CUORICINO
Astropart.\ Phys.\  {\bf 34}, 822 (2011).

\bibitem{EXO(PRL2012)} 
M.~Auger {\it et al.}  [EXO Collaboration],
% Search for Neutrinoless Double-Beta Decay in $^{136}$Xe with EXO-200
Phys.\ Rev.\ Lett. {\bf 109}, 032505 (2012).

\bibitem{KamLANDZ(PRL2013)} 
A.~Gando {\it et al.}  [KamLAND-Zen Collaboration],
% Limit on Neutrinoless $\beta\beta$ Decay of Xe-136 from the First Phase of KamLAND-Zen and Comparison with the Positive Claim in Ge-76,
Phys.\ Rev.\ Lett.\  {\bf 110}, 062502 (2013).

\bibitem{NEMO3(Nd150)}
J.~Argyriades {\it et al.}  [NEMO-3 Collaboration],
% Measurement of the double-beta decay half-life of 150Nd and search for neutrinoless decay modes with the NEMO-3 detector
Phys.\ Rev.\ C {\bf 80}, 032501 (2009).
   
\bibitem{CANDLES}
S.\ Umehara \etal\ [CANDLES Collaboration],
% 
J.\ Phys.\ Conf.\ Ser.\ {\bf 39}, 356 (2006).
  
\bibitem{MAJORANA}
C.E. Aalsetha \etal\ [MAJORANA Collaboration],
% The Majorana Experiment
Nucl. Phys. B (Proc. Suppl.) {\bf217},  44 (2011).

\bibitem{CUORE0}
P. Gorla [CUORE Collaboration],
% SNO+ with Tellurium 
presented at $\nu$Mass 2013, 4-7 February, Milano, Italy;
F.~Alessandria [CUORE Collaboration],
% Sensitivity of CUORE to Neutrinoless Double-Beta Decay
arXiv:1109.0494.

\bibitem{CUORE}
L. Pattavina [CUORE Collaboration],
% Status of the CUORE experiment
J. Phys. Conf. Ser. {\bf447}, 012066 (2013).   
 
\bibitem{SNO+}
S. Biller, [SNO+ Collaboration],
% SNO+ with Tellurium 
presented at TAUP 2013, 3-8 September, Asilomar, CA, USA;
O.~Cremonesi and M.~Pavan,
% Neutrinoless Double Beta Decay Experiments,
arXiv:1310.4692.
  
\bibitem{nEXO}
G. Gratta [EXO Collaboration],
% nEXO 
Intensity Frontier Whitepaper, Mar 2013.

\bibitem{NEXT}
V.~Alvarez {\it et al.}  [NEXT Collaboration],
% The NEXT-100 experiment for neutrinoless double beta decay searches (Conceptual Design Report),
arXiv:1106.3630.

\bibitem{RebelMufson} 
B.\ Rebel and S.\ Mufson,
Astropart. Phys. {\bf 48}, 78 (2013).

\bibitem{LV_DC2} 
J.S.\ D\'iaz \etal,
Phys.\ Lett.\ B {\bf 727}, 412 (2013).

\bibitem{Disentangling} 
B.~Altschul,
% Disentangling Forms of Lorentz Violation With Complementary Clock Comparison Experiments
Phys.\ Rev.\ D {\bf 79}, 061702 (2009);
Y.~Bonder,
% Lorentz violation in a uniform newtonian gravitational field,
Phys.\ Rev.\ D {\bf 88}, 105011 (2013).

\bibitem{DBD_interf} 
S.~Pascoli, M.~Mitra and S.~Wong,
% The Effect of Cancellation in Neutrinoless Double Beta Decay
arXiv:1310.6218.
  
\bibitem{0vbb_m0}
G. Feinberg and M. Goldhaber, 
Proc.\ Nat.\ Acad.\ Sci.\ {\bf 45}, 1301 (1959); 
B. Pontecorvo, 
Phys.\ Lett.\ B {\bf 26}, 630 (1968).


\end{thebibliography}
\end{document}